\documentclass[12pt]{iopart}

\usepackage{iopams}
\usepackage{ulem}
\usepackage[dvips]{graphicx}

\begin{document}

\title{Imaging a single atom in a time-of-flight experiment}
\author{A~Fuhrmanek$^1$, A~M~Lance$^2$, C~Tuchendler$^1$, P~Grangier$^1$, Y~R~P~Sortais$^1$ and A~Browaeys$^1$}
\address{$^1$ Laboratoire Charles Fabry, Institut d'Optique, CNRS, Univ Paris-Sud,
Campus Polytechnique, 2 avenue Augustin Fresnel, RD 128,
91127 Palaiseau Cedex, France}
\address{$^2$ NIST, 100 Bureau Drive, Stop 8424, Gaithersburg, MD 20899-8424
USA}
\ead{yvan.sortais@institutoptique.fr}

\date{\today}

\begin{abstract}
We perform fluorescence imaging of a single $^{87}$Rb atom after its release
from an optical dipole trap. The time-of-flight expansion of
the atomic spatial density distribution is observed by accumulating many single atom images. The position of the atom is revealed with a spatial resolution close to $1~\mu$m by a single photon event, induced by
a short resonant probe. The expansion yields a measure of the temperature of a single atom, which is in very good agreement with the value obtained by an independent measurement based on a release-and-recapture method. The analysis presented in this paper provides a way of calibrating an imaging system useful for experimental studies involving a few atoms confined in a dipole trap.
\end{abstract}


\maketitle

\section{Introduction}\label{section: introduction}

Time-of-flight imaging of ultra-cold atomic gases in expansion is a common
way to study their properties. It provides a direct measurement of
the momentum distribution and is therefore routinely used to extract the
temperature of cold thermal samples~\cite{Lett88}. It can also give
access to spatial density or momentum correlations in atomic ensembles. These features have, for instance,  enabled the
observation of bunching (anti-bunching) with bosonic (fermionic) atoms~\cite{Aspect2007,Jin2005}.
They also enable the study of condensed matter phenomena that emerge when confining matter waves in periodic optical
potentials~\cite{Bloch2008a,Bloch2005}.
Ultimately, one would like to observe the atoms individually in
these mesoscopic systems, not only when the atoms are confined but also when they
move or are released from the trap in order to access out-of-equilibrium properties.
While fluorescence imaging is widely used in experiments to
detect single trapped atoms\,\cite{Schlosser2001,Kuhr01,Weiss2007}, and
sometimes spatially resolve them\,\cite{Sortais2007,Miroschnychenko2006,Bakr09},
fluorescence imaging of freely propagating single atoms has been demonstrated only
recently~\cite{Bucker2009}. In that experiment, cold atoms are released from a trap
and fall under the gravity through a sheet of light, which is imaged on
an intensified CCD camera using efficient collection optics. The presence of an atom is revealed by an individual spot
corresponding to the detection of many fluorescence induced photons.
The detection efficiency of a single atom is close to unity and the spatial
resolution ($\lesssim10~\mu$m) is set by the motion of the atom in the light sheet.

In this paper, we present a complementary approach where
we demonstrate time-of-flight fluorescence
imaging of a single Rb atom in free space, based on single photon detection, with a spatial resolution of $\sim1~\mu$m. A single atom is first trapped
in a microscopic dipole trap and then released in free space where
it evolves with its initial velocity. To detect the atom and locate it with the best accuracy possible, we illuminate it
with a very short pulse of resonant light and collect the fluorescence on an
image intensifier followed by a CCD camera. The presence of the atom is revealed by a
single photon event and, for probe pulses as short as $2~\mu$s, the probability to detect an atom in a single shot of probe light is $4.4\%$.
We repeat the experiment until the spatial distribution of the atom
is reconstructed with a sufficient signal-to-noise ratio; the accumulation
of successive single atom images yields an average result
that exhibits the same features as would a single experiment
with many non-interacting atoms. Average images recorded for
increasing time-of-flights allow us to measure the root mean square (abbreviated rms)
velocity of the atomic expansion, and thus the temperature of a single atom. Our
method allows us to measure temperatures over a wide range.

The analysis of the time-of-flight of a single atom released from our microscopic dipole trap also serves as
a calibration of our imaging system. This calibration will be used in future experiments where we plan to study the
behavior of a cloud of a few tens of cold atoms held in the microscopic trap. In this regime the cloud is very
dense and light scattering of near-resonant light used for diagnostic purposes may exhibit a collective behavior
(see e.g.~\cite{Sokolov2009}). It is therefore important to understand the optical response of the imaging system
in the single atom case to interpret the images in the multi-atom regime where collective effects may come into play. Moreover, because the atoms can be illuminated right after their release from the dipole trap, our method allows us to explore the properties of the momentum distribution of such a gas in the near-field regime.

The paper is organized as follows: in Sect.~\ref{section:experimental_setup}
we describe the experimental setup. In Sect.~\ref{section:requirements}
we explain the requirements to perform single atom time-of-flight imaging.
Section\,\ref{section:experimental_sequence_and_results} describes the experimental
sequence and shows images of a single atom taken by a CCD camera after a
variable time-of-flight.
We extract the temperature using a standard fit
based on an expansion model.
Section~\ref{section:spatial_resolution} details the effects that contribute
to the spatial resolution of our imaging system.
Section~\ref{section:temperature} re-analyzes the data by using a
Monte Carlo simulation taking into account the effects described in
Sect.\,\ref{section:spatial_resolution}. The temperature result is also
compared to an independent measurement based on
a release-and-recapture technique.
Finally, Sect.\,\ref{section:noise} examines the
noise sources in our imaging system.

\section{Experimental setup}\label{section:experimental_setup}

Our experimental setup has been described elsewhere~\cite{Sortais2007}
and is summarized in figure~\ref{figure:Setup}.
Briefly, a single rubidium 87 atom is trapped in a
tight optical dipole trap.
The dipole trap is produced by focusing a laser beam
($\lambda_{\rm trap}=850$\,nm)
down to a spot with waist $w_0=1.1~\mu$m, using a high numerical aperture aspheric lens
($\textrm{NA}=0.5$). The trap depth can be as large as $20$\,mK for a laser power of $80$\,mW.
We use the same lens to collect the fluorescence light ($\lambda_{\rm fluo}=780$\,nm),
which is sent onto an avalanche photodiode (APD) and an image intensifier~\footnote[1]{Model C9016-22MGAAS from Hamamatsu.
The phosphor screen of the intensifier is then imaged onto the
CCD camera through a $1:1$ relay lens.
} followed by a low noise CCD camera~\footnote[2]{Model Pixis
1024 from Princeton Instruments.}
(see below for more details).

In order to image the atom, we illuminate it with probe light, which consists of two counter-propagating beams (to avoid radiation pressure force) in a $\sigma_+ - \sigma_-$ configuration, and is resonant with the
$(5~^2S_{1/2},F=2)$  to $(5~^2P_{3/2},F'=3)$ transition.
The saturation parameter of the probe light is $s=I/I_{\rm{sat}}\sim 1$ for each beam.
We also superimpose repumping light on the probe beams, tuned to the
$(5~^2S_{1/2},F=1)$ to $(5~^2P_{3/2},F'=2)$ transition.

\section{Requirements for time-of-flight imaging of a single atom}\label{section:requirements}

The principle of a time-of-flight experiment
is to measure the position of atoms after a period of free expansion.
From the rms positions of the atoms, one extracts the rms
velocity $\sigma_v$ of the atoms. In our experiment,
we measure the position of the atom by illuminating it with resonant laser light and collecting its fluorescence.
This method requires that the position of  the atom change by less than the resolution of the imaging system
during the light pulse. In our case, the imaging system
is diffraction limited with a resolution
$\sigma_{\rm diff}=0.5~\mu$m. This imposes a pulse duration of $\tau < \sigma_{\rm diff}/\sigma_v$. Typically, for a
rubidium atom at the Doppler temperature ($T_{\rm Doppler}\simeq150~\mu$K),
this yields probe pulses as short as $4~\mu$s. For a
collection efficiency of $\sim1\%$ and a scattering rate
$R\approx\Gamma/2\simeq 2\times10^7$\,s$^{-1}$
($\Gamma$ is the line width of the optical transition),
the number of detected photons per pixel would approach unity in single shot, which is well below the capabilities of our CCD camera.
%This fact imposes that the CCD camera be left open and that a sufficient number of shots be accumulated before reading the CCD register.
%Nevertheless, the presence of stray light due to the trapping laser beams would lead to a dramatic increase of the number of accumulations required to build up an image with good signal-to-noise ratio.

\begin{figure}
\begin{center}
\includegraphics[width=11cm]{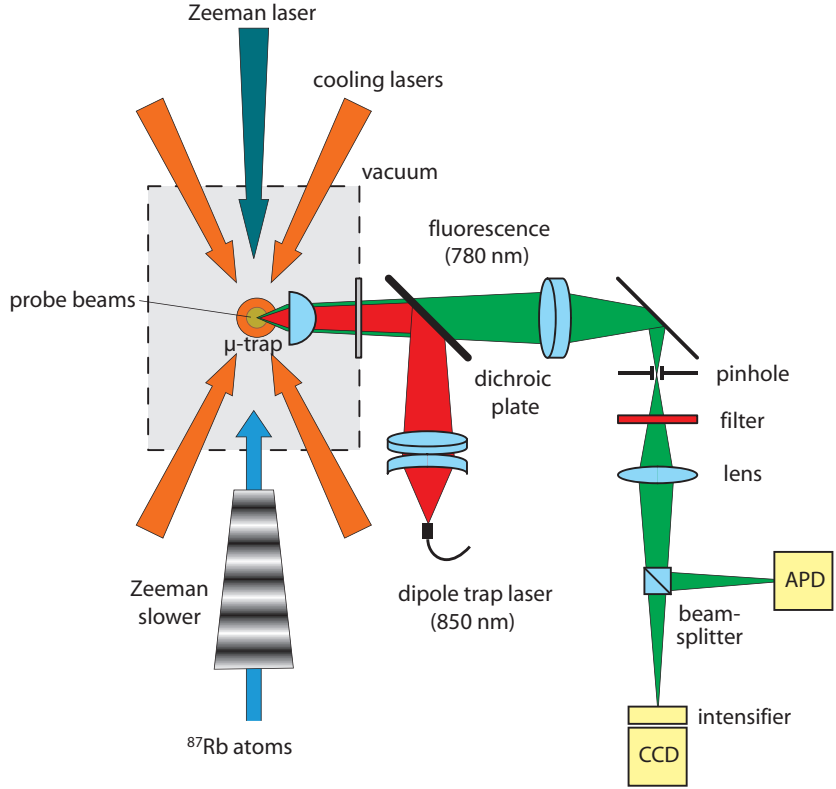}
\caption{(Color Online) Optical setup of our trapping (red) and imaging (green) systems.
A single atom is trapped at the focal point of the aspheric lens and emits
fluorescence photons when illuminated with probe beams (perpendicular to the plane of the figure).
A fraction of the emitted fluorescence is sent to a fiber-coupled avalanche photodiode
(APD) in a photon counting mode, which is used to trigger the time-of-flight sequence.
The remaining fluorescence is also detected by an intensified CCD camera (I-CCD),
with an efficiency $\eta_d=2.2\times 10^{-3}$, taking into account the solid angle of the
aspheric lens ($6.7\%$), the transmission of the optics ($33\%$) and
the measured quantum efficiency of the intensifier photocathode ($\eta_{\rm intensifier}=10\%$).}
\label{figure:Setup}
\end{center}
\end{figure}

We solved this issue by inserting a light intensifier in front of the
CCD camera (see figure~\ref{figure:Setup}). The intensifier acts as a fast shutter (opened during the probe pulse only), and amplifies a single photon event to a level about two orders of magnitude above the noise level of the CCD camera.
%In the absence of any background light, it improves the signal to noise ratio by more than two orders of magnitude.
Using this intensifier, the presence of one atom is revealed by one single photon event (the case of detecting more than one photon emitted by a single atom during the probe pulse is very unlikely).

%suggestion Philippe : \textit{It is therefore needed to accumulate many such shots to get an acceptable image. It should be noticed that if only a CCD camera is used, individual shots cannot be registered and post-processed, because the signal level  is well below the read-out noise of the CCD chip. On the other hand, one cannot simply leave the CCD open and accumulate shots, because of the very  large amount of stray light when reloading the atom : a fast shutter with µs time resolution  is actually needed.
%These problems can be solved by inserting a light intensifier in front of the CCD camera (see Fig. 1). This amplifier acts as a fast shutter, and also amplifies a single photon event above the noise level of the CCD camera. One can  then register single shots, where clearly visible single photon events correspond to the presence of individual atoms.}

\section{Experimental sequence and results}\label{section:experimental_sequence_and_results}

\begin{figure}
\begin{center}
\includegraphics[width = 10cm]{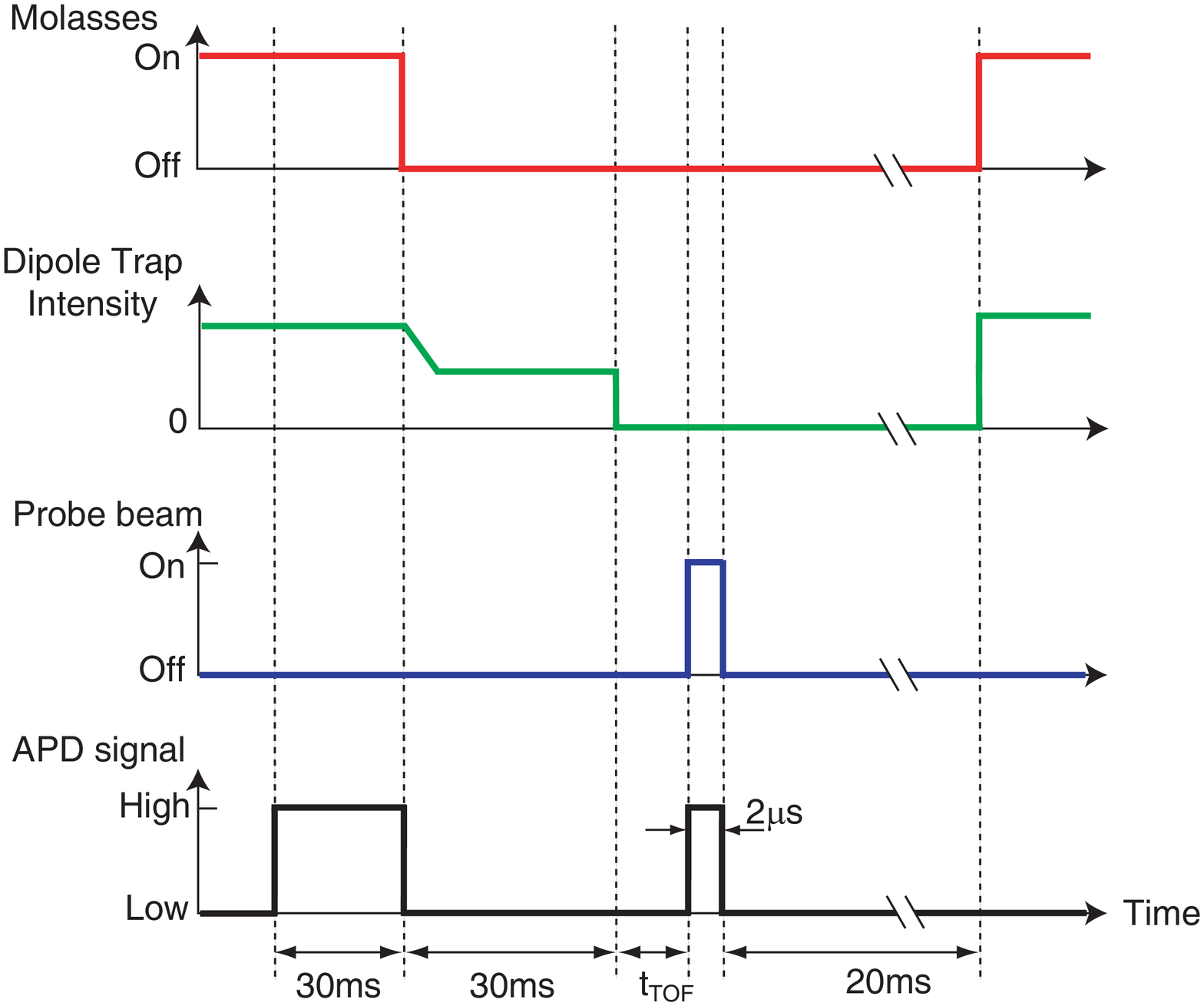}
\caption{(Color Online) Time sequence for a time-of-flight experiment.
Note that the time axis is not drawn to scale. The loading sequence lasts $\sim1$\,s,
while the adiabatic cooling, time-of-flight, and probing sequences are much shorter.}
\label{figure:Experimental_sequence}
\end{center}
\end{figure}

The experimental sequence is summarized in figure~\ref{figure:Experimental_sequence}.
It starts with loading and cooling a single atom in the dipole trap.
Under illumination by the cooling beams, photons scattered by the
trapped single atom are partially collected by the same aspheric lens
and directed towards the APD. This light is used as a trigger signal
for the subsequent time-of-flight sequence. The cooling beams are
switched off immediately upon detection of the atom. The single atom
is kept in the dipole trap for an extra $30$\,ms where the trapped atom
can be further cooled by adiabatically ramping down the trap depth\,\cite{Tuchendler2008}. We also use this $30$\,ms
interval to let the atoms in the molasses spread out, with all cooling
beams having been switched off. This precaution is taken in order to minimize
light scattered by the background molasses during the subsequent probe pulse.

After the single atom is trapped and cooled, the dipole trap is switched off and the single atom time-of-flight
experiment takes place. We let the single atom fly for a variable time $t_{\rm TOF}$ and then illuminate
it by a $2~\mu$s pulse of probe light. At the same time, the intensifier
is switched on for $2~\mu$s and the probe-induced fluorescence is collected by the
intensified CCD camera. The loading sequence is then started again,
in order to prepare for the next time-of-flight experiment.
The acquisition of one image for a given time-of-flight is performed by repeating the experimental
sequence described above, with a cycle rate of $\sim 0.5 - 2$\,s$^{-1}$
and accumulating the total fluorescence light on the CCD. When a
sufficient number of photons have been detected (typically $100$),
the CCD chip is read out and the image is displayed. Note that, for each sequence,
the CCD receives light only during the $2~\mu$s the intensifier is on. In this way the intensifier
also serves as a fast switch, preventing stray light from reaching the CCD during the
cooling and trapping phases.

\begin{figure}
\begin{center}
\includegraphics[width = 11cm]{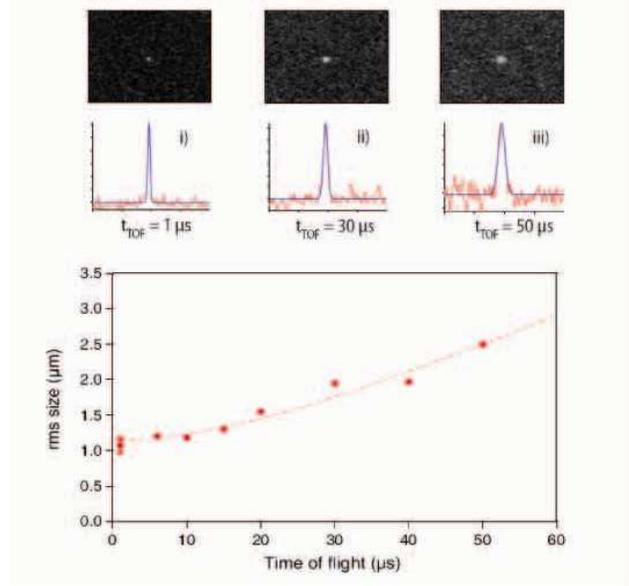}
\caption{(Color Online) Results of a typical single atom time-of-flight experiment.
The atom is released from a trap with depth $U\approx0.08$\,mK.
The rms size of the ``cloud'' is plotted versus the time-of-flight of the single atom
after it is released from the dipole trap. The dashed line is a fit to the data (circles),
using Eq.~(\ref{eqn:TOF_fit}). We show a typical error bar
obtained by repeating the same experiment several times. Insets show
images and associated cross-sections of the data for three particular
time-of-flights. Each image results from the detection of a large number of successively
trapped single atoms. The rms size of the cloud is thus the rms position
of a single atom after a given time-of-flight. Image i) corresponds to
$3400$ sequences and $150$ detected photons, and therefore $150$
detected atoms. Image iii) corresponds to $12000$
accumulations and $520$ detected atoms.}
\label{figure:TOF_Images_Raw_Data}
\end{center}
\end{figure}

Figure~\ref{figure:TOF_Images_Raw_Data} shows typical images taken
for time-of-flights as long as $50~\mu$s. The longer the time-of-flight,
the lower the peak signal, and the larger the number of accumulations required.
For a measured rms size $\sigma\simeq 1~\mu$m (corresponding to the time-of-flight $t_{TOF}=1~\mu$s of image i)), we perform $\sim3400$ sequences, corresponding to $3400$ single trapped
atoms, and detect $150$ photons (this number of photons is extracted from
an independent calibration of the intensifier response to a single photon event).
This means that the probability to detect a single atom in a single realization of
the experiment is $4.4\times10^{-2}$ when using a 2~$\mu$s-probe.

The images are well fitted by a 2D
Gaussian model. Within the error bars the images are isotropic.
We plot the rms size $\sigma$
of the expanding ``cloud" along one axis versus the time-of-flight $t_{\rm TOF}$.
We fit the data shown in figure\,\ref{figure:TOF_Images_Raw_Data} by
the general form
\begin{equation}\label{eqn:TOF_fit}
\sigma(t_{\rm TOF})=\sqrt{\sigma(0)^2+\sigma_v^2 t_{\rm TOF}^2}
\end{equation}
that gives the rms position of a particle after a time-of-flight
$t_{\rm TOF}$ when the initial position
and the velocity are taken from distributions with standard
deviations $\sigma(0)$ and $\sigma_v$. We find $\sigma(0)=1.1(1)~\mu$m
and $\sigma_v=45(2)$\,mm.s$^{-1}$.
The energy distribution of a single atom in the trap being
a thermal Maxwell Boltzmann distribution~\cite{Tuchendler2008},
this translates into a temperature $T=m \sigma_v^2/k_{\rm B}=21(2)~\mu$K
($m$ is the mass of a rubidium atom and $k_{\rm B}$
is the Boltzmann constant).

Let us now compare the result for $\sigma(0)$ to the expected
rms radial position of an atom in equilibrium and trapped in a harmonic potential
with depth $U$ and transverse size $w_0$ (at $1/e^2$), i.e.
\begin{equation}\label{eqn:size_in_trap}
\sigma_\perp(0)=\sqrt{\frac{k_B T}{m \omega_\perp^2}}\ ,
\end{equation}
where $\omega_\perp$ is the radial
oscillation frequency of the atom in the trap. With  $\omega_\perp~\approx~2\pi\times 26$\,kHz
and $T = 20\ \mu$K, we find $\sigma_\perp(0)=0.3~\mu$m,
below the diffraction limit of the imaging system. Taking the latter into
account, we should thus expect a rms size of $0.6~\mu$m at null
time-of-flight, i.e. a factor $1.8$ below the actual data.

\section{Spatial resolution of our imaging system}\label{section:spatial_resolution}

In order to understand the size at $t_{\rm TOF}=0$,
we investigated experimentally the effects that contribute to the loss in resolution
of our imaging system and lead to the measured value $\sigma(0)$.
These effects are listed in table~\ref{table:Imaging_budget}
and sum up quadratically to yield a value of $1.1~\mu$m, in agreement with the
measure of $\sigma(0)$ obtained in
Sect.~\ref{section:experimental_sequence_and_results}.

\begin{table}
\begin{center}
\begin{tabular}{|l|c|}
    \hline
    Effect & rms size ($\mu$m) \\
    \hline
    Intensifier & $0.9(2)$ \\
    Diffraction ($\sigma_{\rm diff}$) & $0.5(1)$ \\
    Atomic thermal distribution ($\sigma_\perp(0)$) & $0.3$ \\
    Depth of focus ($\sigma_\parallel(0)$) & 0.1 \\
    Atomic displacement during $\tau$ ($\sigma_{\tau,\rm thermal}$) & $0.04$ \\
    Atomic random walk ($\sigma_{\tau,\rm scatter}$) & $0.02$ \\
    \hline
    Quadratic sum & $1.1(2)$ \\
    \hline
  \end{tabular}
  \end{center}
\caption{Spatial resolution budget of our system, when the light source is a
single atom with temperature $T=20~\mu$K illuminated by a probe pulse with
 duration $\tau=2~\mu$s (see text). The rms size of the global response is the
 quadratic sum of the different rms contributions.}
 \label{table:Imaging_budget}
\end{table}

The dominant contribution comes from the loss of resolution of the imaging
system due to the intensifier. This contribution was measured by imaging
an object with a sharp edge and characterizing the blurred edge in the image obtained.
The second largest contribution comes from the
diffraction limit of the imaging optics, which is due to the numerical aperture of
the aspheric lens, and was tested by removing the intensifier and illuminating
a trapped atom for $100$\,ms. In this case, the atom acts as a point source for the
imaging system and the associated response on the CCD is well fitted by a
Gaussian shape with a size $\sigma_{\rm diff}=0.5(1)~\mu$m.

\begin{figure}
\begin{center}
\includegraphics[width = 12.5cm]{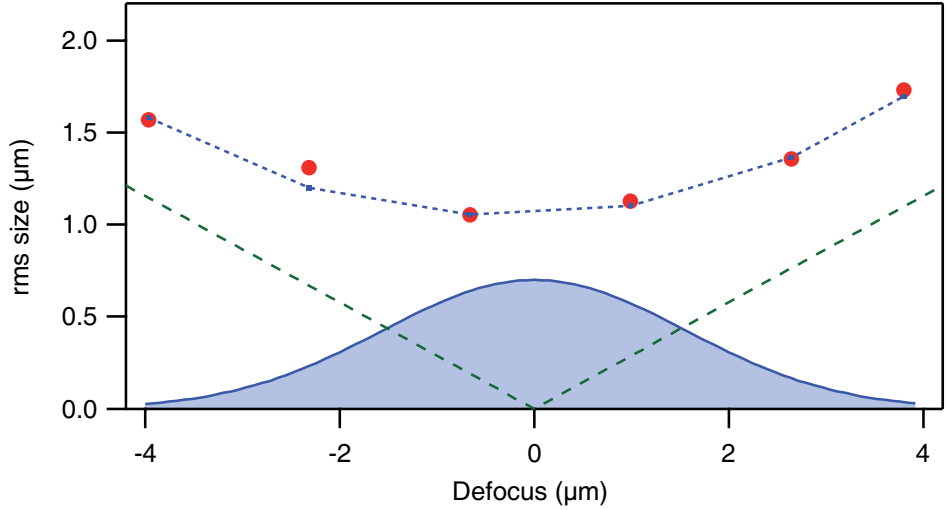}
\caption{(Color Online) rms size of the image of a single trapped atom for various
positions of the trap along the optical axis of the imaging system. For small
values of the defocus, the image size is given essentially by the response of the intensifier.
For large values of the defocus, the main contribution comes from the depth
of focus, which scales linearly with defocus according to equation~(\ref{eqn:defocus})
(dashed line). Our data (red circles),
which agree well with a Monte Carlo simulation (blue dotted line), result from
the convolution of both the optical response of the imaging system
and the longitudinal profile of the thermal distribution (blue filled curve) with a width $\sigma_\parallel(0)=1.6~\mu$m.}
\label{figure:Depth_of_focus}
\end{center}
\end{figure}

The thermal distribution contributes for $0.3~\mu$m due to the
transverse size of the distribution $\sigma_\perp(0)$, and $0.1~\mu$m
due to the effect of depth of focus associated to the longitudinal size of the
distribution $\sigma_\parallel(0)$. We measured this effect of the
depth of focus by imaging a single trapped atom for various positions of the
trap along the optical axis of the imaging system.
Fig.~\ref{figure:Depth_of_focus} shows the rms size of a Gaussian fit to the
data, although for large values of the defocus
$\delta z$ they slightly deviate from a Gaussian.
The results tend asymptotically to the expected
 rms value of a disc with uniform intensity
\begin{equation}\label{eqn:defocus}
\sigma_{\rm defocus}=\frac{1}{2}\, \delta z \times \tan \alpha
\end{equation}
where $\alpha$ is related to the numerical aperture by $\sin\alpha = \rm NA$.

Finally, we analyze the contribution of the movement of the atom during the probe pulse.
Firstly, the photons scattering by the probe induces a random walk of the atom,
leading to a rms position in the plane perpendicular to the probe beam
\begin{equation}\label{eqn:scatter}
\sigma_{\tau,\rm scatter}=\frac{1}{3}v_{\rm rec}~\sqrt{R}~\tau^{3/2}
\end{equation}
where $v_{\rm rec}$ is the recoil velocity, $R$ is the
spontaneous emission rate, and $\tau$ is the duration of the probe pulse\,\cite{Joffe1993}.
Secondly, the atom moves during the probe pulse due to the thermal velocity.
An analytical calculation of the associated rms displacement yields
\begin{equation}\label{eqn:thermal}
\sigma_{\tau,\rm thermal}=\sigma_v~\tau/\sqrt{3}
\end{equation}

\begin{figure}
\begin{center}
\includegraphics[width = 12.5cm]{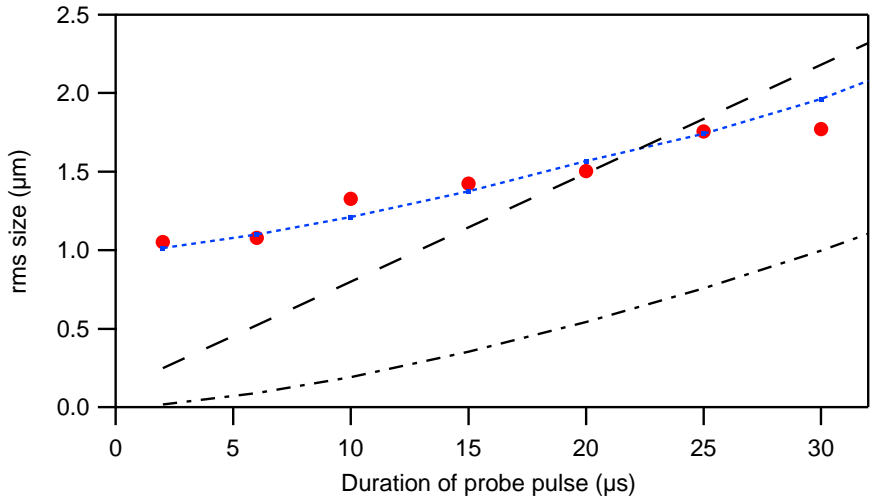}
\caption{(Color Online) Effect on the measured spatial distribution of a single atom as a result of its movement during
a probe pulse of variable length, after $1~\mu$s time of flight. The temperature of the atom is $150~\mu$K.
Data (red circles) are in good agreement with the Monte Carlo simulation
results (blue dotted line). At low pulse durations, the image size is
limited by the intensifier response. Also shown : analytical rms
contributions of equation\,(\ref{eqn:scatter}) (dash dotted line) and of equation\,(\ref{eqn:thermal})
(dashed line). The fact that the contribution of equation\,(\ref{eqn:thermal}) is larger than both experimental and simulated results for long probe durations is due to the significant deviation of the distributions from a Gaussian model.}
\label{figure:Probe_duration}
\end{center}
\end{figure}

Both contributions (\ref{eqn:scatter}) and (\ref{eqn:thermal})  broaden
the image of a single atom when the duration of the probe $\tau$ is increased.
We tested this effect by increasing $\tau$ up to $30~\mu$s, as shown in
figure~\ref{figure:Probe_duration}. Although negligible for $2~\mu$s
probe pulses and atoms at $150~\mu$K (as is the case in figure~\ref{figure:Probe_duration}),
this effect alone would be comparable to the intensifier response if we were using pulses as long as
$20~\mu$s in the perspective of scattering more photons per shot and thus detect single atoms with a larger efficiency.

We confirmed the analysis above by a Monte Carlo simulation that takes into
account all the effects mentioned above. It reproduces accurately our experimental data (see figure\,\ref{figure:Depth_of_focus} and figure\,\ref{figure:Probe_duration}).
Here, we note that the simulation indicates a significant deviation
from a Gaussian shape for long probe durations or large values of the defocus.
Because of the presence of noise in our imaging system, we did not consistently
observe significant deviations on the real images and could not calculate any
reliable value for the rms size of the images. We thus fitted our images with a Gaussian model and
compared it to a Gaussian fit of our simulation. The discrepancy between the
analytical expression (\ref{eqn:thermal}) and
the results in figure\,\ref{figure:Probe_duration}
is an indication of the error that we make by doing so. Note also
that we have not included in the model the potential effect of the
cooling  of the atom by the counter-propagating probe beams.

\section{Temperature results and comparison to release-and-recapture experiments}\label{section:temperature}

We now come back to the temperature result obtained by fitting
the data using equation\,(\ref{eqn:TOF_fit}). Among all the effects
degrading the resolution, the depth of focus is the only one that
varies with the time-of-flight as the atom can fly in the direction parallel to
the optical axis. Therefore, equation\,(\ref{eqn:TOF_fit})
is not strictly valid in our case. We now use the Monte Carlo simulation
mentioned above to fit the data shown in figure\,\ref{figure:TOF_Images_Raw_Data}.
The starting point of this simulation is a thermal distribution with a temperature
that we adjust in order to reproduce the data. We find $T=20(2)~\mu$K.
Not surprisingly, this result is in good agreement with the rough analysis mentioned in Sect.\,\ref{section:experimental_sequence_and_results}, since the effect of the depth
of focus is small.

To cross check the temperature measurement, we
use an independent method based on a release-and-recapture
technique described in detail in reference~\cite{Tuchendler2008}. This method
uses an APD to detect the presence or the absence of the atom
in the trap after release for a variable time. Figure\,\ref{figure:Rel_Rec}
shows the results obtained by the release-and-recapture method.
A fit to the data yields a temperature $T=19(2)~\mu$K in
good agreement with the results of the time-of-flight method.

\begin{figure}
\begin{center}
\includegraphics[width = 12.5cm]{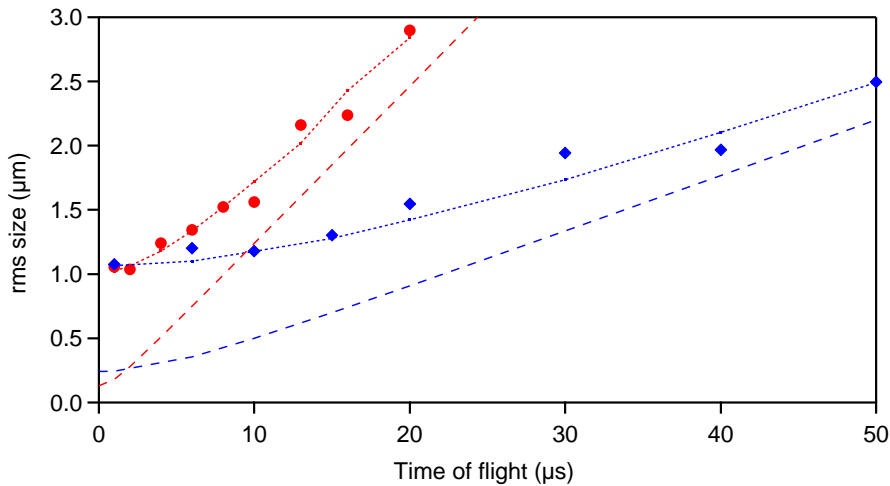}
\caption{(Color Online) Time-of-flight measurements of a single atom for two different temperatures : $T=20(2)~\mu$K (blue diamonds) and
$T=150(16)~\mu$K (red circles). The simulated results are also shown (blue and red dotted lines respectively). The dashed lines
(respectively blue and red) show the results of a simple model neglecting all broadening effects except for the transverse dimension
of the thermal distribution, $\sigma_\perp(0)$.}
\label{figure:TOF}
\end{center}
\end{figure}

The two methods presented above can be extended to higher
temperatures. For example, figure~\ref{figure:TOF} and
figure~\ref{figure:Rel_Rec} show results obtained for a temperature
$T\sim150~\mu$K, achieved by leaving the dipole trap depth unchanged after loading it with a single atom. Our
detection method is therefore applicable over a large range of temperatures
with no anticipated limitation in the low temperature range.

\begin{figure}
\begin{center}
\includegraphics[width = 12.5cm]{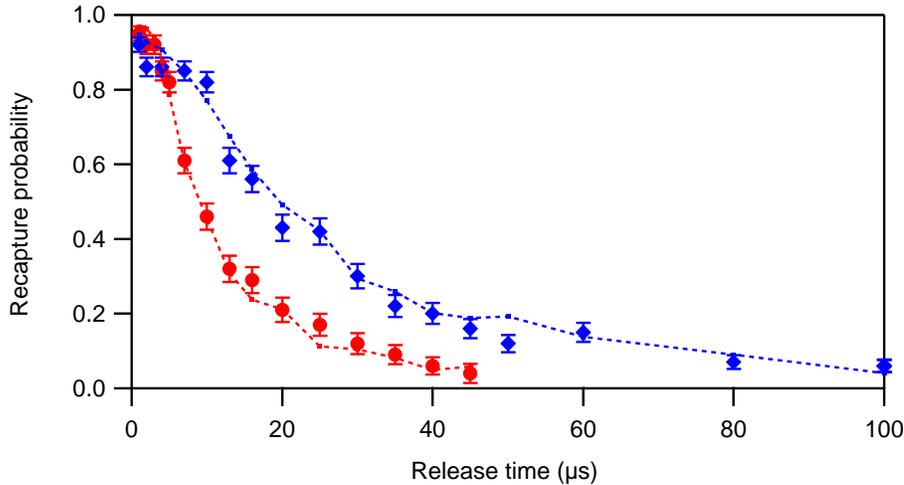}
\caption{(Color Online) Release-and-recapture measurements of a single atom
for the same loading and cooling parameters as in figure\,\ref{figure:TOF}.
The fit (dotted lines) yields $T=19(2)~\mu$K and $149(15)~\mu$K (data points are in diamonds and circles, respectively).}
\label{figure:Rel_Rec}
\end{center}
\end{figure}

\section{Analysis of the noise of the imaging system}\label{section:noise}

\begin{figure}
\begin{center}
\includegraphics[width = 12.5cm]{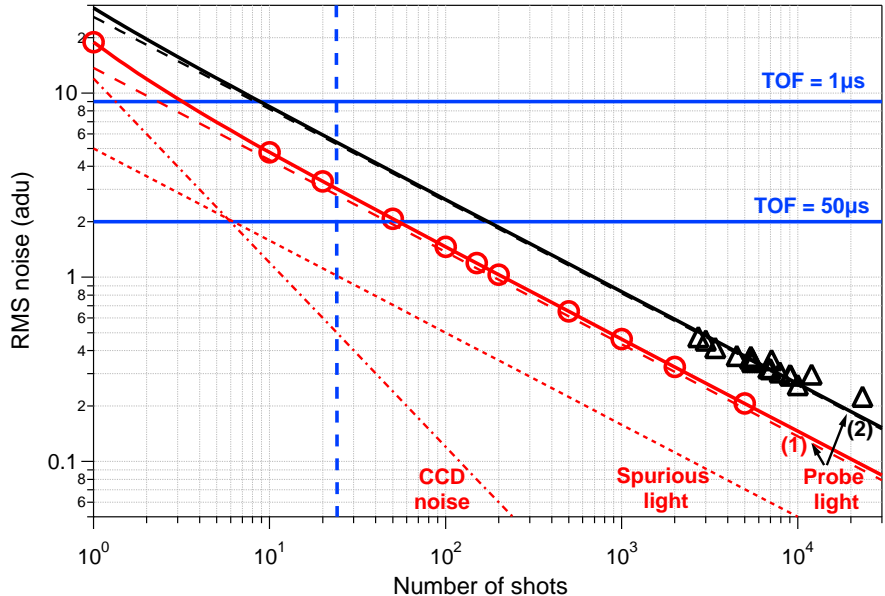}
\caption{(Color Online) Noise contributions, normalized to one shot, of our imaging system, versus the number of shots performed to acquire an image. The total rms noise (solid lines) is the quadratic sum of three contributions that were measured independently : the CCD read-out noise (red dash dotted line) and the background contributions due to spurious light (red dotted line) and probe light (dashed lines). The latter is larger when the atomic beam is switched on (case (2): black dashed line) than when it is off (case (1): red dashed line), due to the scattering by atoms of the atomic beam intersecting the probe beam. In both cases, the quadratic sum of the three contributions is in good agreement with the measured rms values of the total noise  (black triangles and red circles, respectively). The peak signal, normalized to one shot (blue horizontal lines) is shown for two values of the time-of-flight, $t_{\textrm{\scriptsize{TOF}}}=1~\mu$s and $t_{\textrm{\scriptsize{TOF}}}=50~\mu$s, corresponding to images shown in figure~\ref{figure:TOF_Images_Raw_Data}(i) and (iii). The vertical dashed line indicates the minimum number of shots required to detect one atom.}
\label{figure:Noise_imaging_system}
\end{center}
\end{figure}

We now address the issue of the noise of our imaging system.
The peak signal in the time-of-flight image shown in
figure~\ref{figure:TOF_Images_Raw_Data}(i) is $30600$
adu in $1$ pixel~\footnote[3]{$1$\,adu (analog-to-digital unit) corresponds to
the digitization step of the analog signal acquired by the CCD camera.
For the measurements presented in this
paper, the electron-to-adu conversion factor was $1~\textrm{e}^-/\textrm{adu}$.}
 and corresponds to the detection of $\sim150$
single atoms after $3400$ shots of $2~\mu$s probe pulses.
Normalized to one shot, the mean peak signal is thus $9$\,adu in $1$ pixel.
This should be compared to the background noise,
which results from three contributions shown in
figure~\ref{figure:Noise_imaging_system}: read-out noise
from the CCD camera~\footnote[4]{The dark count of the CCD being
$\sim 0.001$ adu/pixel/s is negligible for the parameters of our experiment.}, a background noise contribution
from the probe light, and a background noise
contribution from spurious light (other than probe light).

We have measured the read-out noise of the CCD camera and
found $12$ adu in one image. This noise is independent
of the number of shots $N_{\rm seq}$
performed to acquire the image, since the CCD is read out only once after
the probe pulses have illuminated the atom and the associated scattered
light has fallen on the CCD. Normalized to one shot,
the read-out noise of the CCD thus scales as $12~\rm{adu}/N_{\rm seq}$.
By contrast, the contributions, per shot, of the probe light and spurious light,
scale as $1/\sqrt{N_{\rm seq}}$. The three contributions were
measured independently and add up quadratically to yield
the data shown in figure\,\ref{figure:Noise_imaging_system}.
The signal to noise ratio is by far limited by the probe
light contribution, which is due in part to scattering by
atoms of the Rb beam intersecting the trapping region, and in
part to scattering by the optics mounts under vacuum.

Figure~\ref{figure:Noise_imaging_system} allows us to extract the
number of sequences necessary to reach a given signal-to-noise ratio.
As explained at the end of Sect.~\ref{section:experimental_sequence_and_results},
the probability to detect one photon (and therefore one atom)
in single shot is $4.4\times10^{-2}$ using a $2~\mu$s-duration probe,
meaning that $23$ shots are necessary to detect on average one photon.
For a time-of-flight image to be correctly fitted, we have found that
we need typically $100$ detected photons, which implies $2300$ sequences.
For instance, in the case of the image shown in figure~\ref{figure:TOF_Images_Raw_Data}(i),
the signal to noise ratio is $\sim20$, while it is $\sim 9$ for
figure~\ref{figure:TOF_Images_Raw_Data}(iii).

\section{Conclusion}\label{section:conclusion}
In conclusion, we have demonstrated fluorescence imaging of a single atom in free flight by accumulating many images
containing a single photon event corresponding to a single atom. We used this time-of-flight technique
to measure the temperature of the atom after release from a microscopic optical dipole trap. This temperature measurement was confirmed by an
independent method based on a release-and-recapture technique. The large numerical aperture of our imaging system
and the extreme confinement of the atoms in the trap allow a high spatial resolution on the order of $\sim1~\mu$m. The low noise level of our imaging system yields images showing $\sim 150$ atoms with a very good signal to noise ratio ($\sim20$). These measurements have been performed in conditions where the atomic motion during the probe pulse can be completely neglected (see table\,\ref{table:Imaging_budget}). We thus obtained a very accurate characterization of the optical performance of our system. In a next step, it will be possible to increase the single atom detection efficiency, by increasing the probe pulse duration (to the expense of a somehow reduced spatial resolution and signal-to-noise ratio). Finally, the measurements and the analysis presented in this work provide a calibration of our imaging system for future time-of-flight experiments where many atoms are confined in a microscopic dipole trap, and where interactions may play a central role.

\ack{We acknowledge support from IARPA, the Institut Francilien de Recherche sur les Atomes Froids (IFRAF), PPF ``Manipulation d'atomes froids  par des lasers de puissance''  and the European Union through the Integrated Project SCALA. A~M~Lance was supported by an E.U. Marie Curie fellowship and A.~Fuhrmanek is partially supported by the DAAD Doktorandenstipendium}.

\end{document}